# Simulation and Experimental Validation of Optical Camera Communication

Srivathsan Chakaravarthi Narasimman
*School of Electrical and Electronics Engineering,*
Nanyang Technological University, Singapore
srivaths003@e.ntu.edu.sg

Arokiaswami Alphones
*School of Electrical and Electronics Engineering,*
Nanyang Technological University, Singapore
EAlphones@ntu.edu.sg

*Abstract*— Visible light communication (VLC) with photo detectors (PDs) have been widely investigated, similar tools for optical camera communication (OCC) with complementary metal oxide semiconductor (CMOS) sensors are lacking in this regard. Camera based VLC systems have much lower data rates owing to camera exposure times. Among the few extant OCC simulation tools, none allow for simulation of images when exposure time is greater than the signal period. An accurate simulation of the OCC system can be used to improve the data rate and quality of performance. We propose a simple simulation technique for OCC which allows to test for system performance at frequencies beyond the camera shutter speed. This will allow much needed data rate improvement by operating at the actual frequency a decoding algorithm ceases detection instead of the exposure limit used now. We have tested the accuracy of simulation by comparing the detection success rates of simulated images with experimental images. The proposed simulation technique was shown to be accurate through experimental validation for two different cameras.

*Keywords— Visible light communication, optical camera communication, simulation, decoding, CMOS sensor, image processing.*

## I. INTRODUCTION

The existing standards for VLC systems ranging from IEEE 802.15.7 to 802.11bb prioritize the use of PDs over cameras, owing to high speeds achieved by the former. However, the ubiquity of cameras provides important use cases for camera-based VLC such as indoor positioning [1], [2] and vehicular communication [3]. Cheap CMOS sensors can be used for unidirectional communication to facilitate indoor positioning. Coarse transmitter localization can be achieved using radio fingerprinting [4] or magnetic field strength-based techniques [5], but they are not as accurate as camera-based VLC. To identify each transmitter uniquely, the length of code being transmitted has to be high, which requires a high data rate. Simulation tools are generally used to identify the maximum frequency of operation and hence decide the maximum number of bits in the code. While PD based VLC has several tools to identify this, camera-based VLC does not.

Current OCC simulation techniques are not as accurate as in the case of PDs, owing to the different processing pipelines employed by CMOS cameras. A radiometric approach with a complete image processing pipeline was produced to test camera performance [6] but the rolling shutter effect of CMOS cameras was not included. A Lambertian model for the transmitter was used to incorporate distance into the simulation [7] and transmitter illumination data was used in Blender to generate photorealistic images of lights [8] but both lack OCC capabilities. CamComSim [9] employs a Markov-modulated Bernoulli process to simulate a network and produce probability of success but it does not generate an image to facilitate decoding algorithm testing. OCC simulation has been performed using DC gain of each pixel in the area of view of the camera [10] and using photometric properties [11] but both do not outline operation beyond shutter speed with the former adhering to the Nyquist rate. Though there are simulation techniques available for OCC, they do not work for frequencies beyond the shutter speed.

Most camera-based VLC simulation techniques use physical principles such as radiation or photometry to determine equations for performance metrics such as signal to interference plus noise ratio or maximum bit rate directly which leads to the rigidity of these techniques. We seek to address this issue using a simple weighted average of expected light intensities over individual exposure periods allowing for the simulation of images close to reality. To test the accuracy of the proposed simulation technique, simulated images were decoded using a commonly used thresholding technique [12] and the results were compared with experimental data.

The main contributions of this work are

• A simple technique to simulate OCC at any signal frequency irrespective of exposure time.

• Experimental validation of simulation for two different cameras.

## II. METHODOLOGY

The process outline is delineated in the Fig. 1, where the transmitter is an LED panel light. A 10 bit code was chosen since it will allow for 1024 unique variations which can be assigned to as many lights in the case of transmitter localization for indoor positioning. This code is then encoded using differential Manchester encoding to limit the run length of same bits since this will cause noticeable flicker at lower frequencies. We have used on off keying (OOK) modulation, which is the most commonly used modulation technique for OCC [10]. Since image processing is computationally intensive, this simple modulation scheme allows for detection on lower end smartphones. By capturing

This Research is supported by the RIE2020 Industry Alignment Fund - Industry Collaboration Projects Funding Initiative (Award No. I1801E0020) administered by the Agency for Science, Technology and Research (A*STAR), as well as cash and in-kind contribution from Surbana Jurong Pte Ltd.

images of the LED panel through a camera the transmitted data is received. To simulate this received image, the camera parameters such as the exposure time, focal length and aperture along with the transmitter properties such as the area of the panel and luminous intensity are used. Samples for the simulated and actual image for the same parameters are shown in the Fig. 1, which look similar. While we can compare the images directly for similarity metrics, it does not tell us if the simulated image will perform the same as the actual image on detection algorithms which is the main use of simulation. Hence, we perform image processing to get the transmitted string which is then decoded to obtain the transmitted code. We compare the received codes with the transmitted code to find the number of correct bits which is the success rate of transmission.

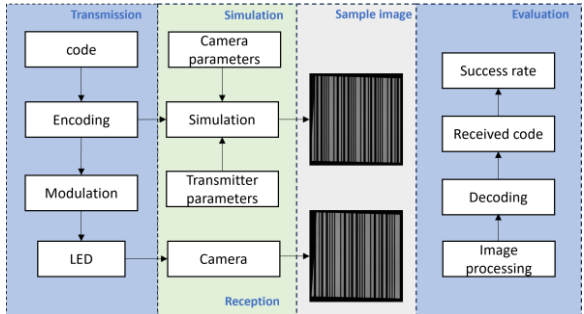

*Fig 1 Process Outline*

### A. Experimental Setup

To test for the robustness of the proposed simulation scheme, we collected images on a phone and tablet front camera at different distances from the transmitter and at different switching frequencies. We use the front camera on both these devices since it allows the user to look at the screen of these devices for navigation when using the lights above them for positioning. The Redmi Note 9 Pro and Galaxy tab S7 front cameras are the phone and tablet with the corresponding readout times being 8 and 13 microseconds. The code to be transmitted was encoded on an Arduino Uno and an n channel MOSFET switch was used to switch the light on and off according to the encoded bit string at switching frequency from 2 to 20kHz in 2kHz increments. To facilitate ease of data capture we placed the light on the floor and the receiver at distances ranging from 60 cm to 200 cm in 20 cm increments. For each frequency, exposure time and distance, five images were captured.

### B. Simulation Technique

The image generation pipeline for CMOS sensors is the same across the current simulation techniques for OCC. The amount incident photons on the sensor is determined by the luminous intensity of the transmitter along with the lens aperture and exposure time of the camera. These photons are converted to electrons based on the quantum efficiency of the sensor and the resulting voltage is amplified based on the ISO speed of the camera. This is then digitized and converted to a pixel value between 0 and 255 through gamma encoding. Each camera processes these pixels through a unique image processing pipeline which is not revealed to the user, following which the final image is generated. The general pixel value determination for when the light is on or off is performed as outlined in [11] and detailed below as $PV_{max}$ and $PV_{min}$ respectively.

$$PV_{max} = 118\left(\frac{S \times t}{K \times N^2}\left(L_v + E_v \frac{R}{\pi}\right)\right)^{1/\gamma}$$

where S is the ISO speed of the camera set to be 100 in our experiments, t is the exposure time, N is the lens aperture, $L_v$ is the luminous intensity of the transmitter which was 3600 cd/m2 in our experiments, $E_v$ is the external illuminance measured to be 290 lux, R is the reflectance assumed to be 40% as per [13], K and γ are constants assumed to be 12.5 and 2.22 as per [11].

$$PV_{min} = 118\left(\frac{S \times t}{K \times N^2} \times \left(E_v \frac{R}{\pi}\right)\right)^{1/\gamma}$$

In a CMOS sensor with rolling shutter, each column is exposed individually for the exposure time. Each column aggregates the amount of light over this time period. When the switching period is less than the exposure time, the light will be on and off within a single exposure period. We calculated the simulated pixel value for each column $PV_{sim}$ as the weighted average of $PV_{max}$ and $PV_{min}$ using the duration for which the light is on and off as the weights. The formula for which is as follows

$$PV_{sim}(i) = \frac{PV_{max} \times t_{on}(i) + PV_{min} \times t_{off}(i)}{t}$$

where $t_{on}(i)$ is the duration for which the light was on when column i was exposed and $t_{off}(i)$ is the corresponding duration when the light was off. There is a small delay between when each column is exposed called the readout time, which is much smaller than the lowest possible exposure time for the camera. Therefore, there is a significant overlap in the states for consecutive columns. This allows us to observe the light states even when the exposure time is greater than the switching period. Since we do not know the exact image processing steps used to arrive at the final image, we employed a commonly used contrast enhancement technique called histogram equalization to improve detection. The pixel value calculation uses transmitter and receiver properties but does not take the distance between them into consideration.

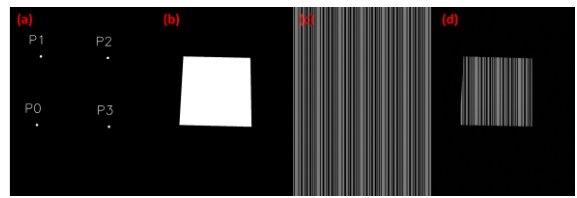

*Fig 2 (a) projected light corners (b) mask of the light area (c) calculated pixel value (d) simulated image*

The larger the area covered by the transmitter in the image better the chances of detection. Using the intrinsic camera properties, we projected known coordinates of the LED corners to the image plane, which are labelled in Fig.2(a). These corners are then joined to form a mask of the image area covered by the transmitter shown in Fig. 2(b). The string of pixel values after histogram equalization is shown in Fig. 2(c), where the light covers the entire image area. By masking the area of interest, we arrive at the final simulated image shown in Fig. 2(d).

## C. Decoding Technique

To test the accuracy of simulation, we propose to compare the detection success rate of simulated and experimental values. We identify the area of the image covered by the transmitter through image processing. We determine the brightest pixels in the image covering five percent of the area, which gives us some of the pixels in the transmitter image. If there are multiple contours the overlapping area is used to grow the contour until only one contour remains. A bounding rectangle is constructed over this contour which is sliced from the original image as shown in the Fig. 1. The columns within this area are averaged to obtain a signal. Otsu's thresholding is used to binarize the image. A histogram of the run length ones and zeros is constructed from which once again Otsu's thresholding is used to determine the average run length for ones and zeros. We used a header with three continuous ones, since the code can never have that owing to differential Manchester encoding. The first two instances of the header are used to separate the code of interest, which is then decoded to get the received code. Since the data rates achieved by OCC are much lower than VLC and the difference between a single transmitted and received code is the metric of importance, we report the success rate as a percentage instead of the bit error rate, which is defined as follows.

$$SR = \frac{\#B_c}{\#I \times \#B_t} \times 100$$

where $\#B_c$ is number of correct bits in the received code, $\#I$ is the number of images and $\#B_t$ is the total number of bits in the code.

## III. RESULTS

### A. Simulated Images

To show that the proposed simulation technique is accurate and better than extant techniques, we compared it to a state of the art (SOTA) simulation technique. The photometry-based simulation technique in [11] is considered SOTA since it defines simulation until the exposure time is less than the switching period while others define simulation when the exposure time is less than half of it. The SOTA defines the complete band and transition band lengths to determine the sequence of pixel values for all columns.

$$h_c = \frac{t_{LED} - t}{t_r} \qquad h_t = \frac{t}{t_r}$$

where $h_c$ is the number of columns that will be at the zero or one state, $h_t$ is the number of columns when moving from one state to another, $t_{LED}$ is the switching period of the LED and $t_r$ is the readout time of the camera. When the exposure time is equal to the switching period the complete band becomes zero as per this technique.

The pixel values when the exposure time is less than the switching period is shown in Fig. 3, where the three plots refer to the SOTA, the proposed technique and the experimental value for 10 kHz signal when the image was taken with an exposure time of 68µs 120 cm from the transmitter. The SOTA pixel values are lower than the other two since contrast enhancement was not performed. The logic used by SOTA to determine complete and transition band lengths is shown to provide different pattern compared to the experimental value even when the exposure time is less than the switching period.

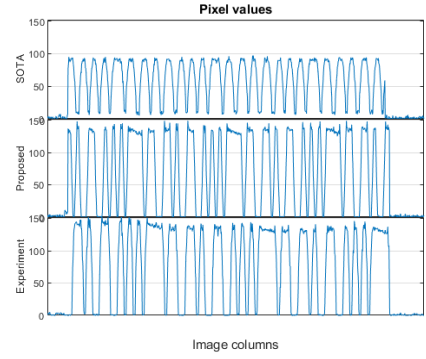

Fig 3 Pixel values when the exposure time is less than the switching period

Since SOTA is not defined when exposure time is greater than the switching period, the pixel values for the proposed simulation technique and experimental values for are reported in Fig. 4. The pixel values of the proposed technique are marginally higher than the experimental values and the patterns are slightly different from each other. This can be due to the different image processing techniques used, but the overall difference between the two similar consecutive bits and one bit is apparent from both. Here the image was captured at the same settings with an exposure time of 136µs. As the exposure time increases the contrast between single ones and zeros reduces eventually changing the sequence of bits rendering detection impossible.

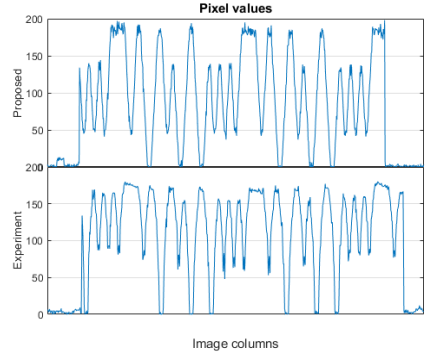

Fig 4 Pixel values when the exposure time is greater than the switching period

### B. Experimental Validation

The detection success rate is used to determine the accuracy of simulation and to ascertain the frequency at which detection stops. We have used two exposure times for both the devices tested, with one being the lowest possible time for that camera and twice that value. The detection results for the phone are shown in Fig. 5, where the experimental success rate for both the exposure times is similar to the simulated values. The experimental success rate for both exposure times at 8kHz beyond 180cm is lower than the simulated value but this is just a difference of two data points which could be chalked up to two headers not being observed for the experimental images. For the higher exposure time though the exact success rate is not observed, the technique provides a good indication of when detection ceases. This happens when the switching period is nearly half the exposure time at 14 kHz.

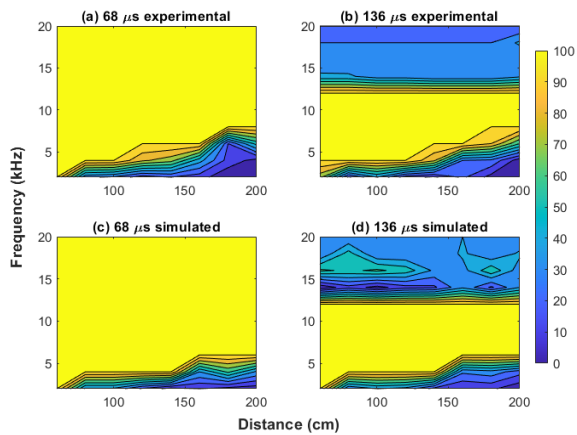

Fig 5 Detection success rate comparison for phone camera.

The success rates for the tablet camera are outlined in Fig. 6, where once again the experimental success rates are similar to the simulated values for both exposure times. The experimental success rate is lower than the simulated value at 2 kHz for both exposure times. We can determine that this frequency is low to accommodate an entire 10 bit sequence in the image. The detection ceases when the switching period is slightly greater than half the exposure time at 16 kHz. Thus, the proposed technique provides the ability to test detection techniques and to determine the ideal number of bits and switching frequency for a given exposure time.

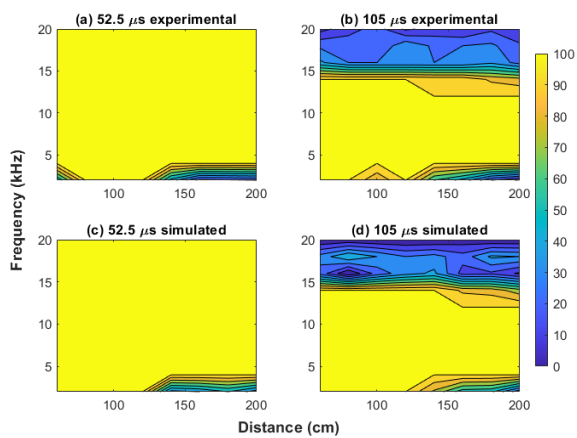

Fig 6 Detection success rate comparison for tablet camera.

## IV. Conclusion

A simple simulation technique for camera-based VLC at any signal frequency irrespective of exposure time was proposed. The simulated images were compared with experimental images. The accuracy of the simulation technique was tested using a simple detection technique to compare detection success rates of simulated with actual images. Thorough experimental validation was conducted using two different devices at two exposure times for a range of switching frequencies and distances. The simulated and experimental success rates were shown to be similar up to the switching frequency where detection was no longer possible for both the devices. Different modulation schemes, coding techniques and detection methods will be further explored.